\documentclass[12pt]{article}
\usepackage{epsf}

\begin{document}
\newcommand{\be}{\begin{equation}}
\newcommand{\ee}{\end{equation}}
\newcommand{\bel}[1]{\begin{equation}\label{#1}}
\newcommand{\bea}{\begin{eqnarray}}
\newcommand{\eea}{\end{eqnarray}}
\newcommand{\ba}{\begin{array}}
\newcommand{\ea}{\end{array}}
\newcommand{\bra}[1]{\mbox{$\langle \, {#1}\, |$}}
\newcommand{\ket}[1]{\mbox{$| \, {#1}\, \rangle$}}
\newcommand{\exval}[1]{\mbox{$\langle \, {#1}\, \rangle$}}

\title{Symmetry effects and equivalences in lattice 
models of hydrophobic interaction}
\author{
G.M. Sch\"{u}tz$^{(1)}$\footnote{Permanent address: Institut f\"ur 
Festk\"orperforschung, Forschungszentrum J\"ulich,
D-52425 J\"ulich, Germany}, I. Ispolatov$^{(1)}$, G.T. Barkema$^{(2)}$\\
and B. Widom$^{(1)}$\\
{\em \small $(1)$ Department of Chemistry and Chemical Biology,}\\ 
{\em \small Baker Laboratory, Cornell University, Ithaca, NY 14853-1301, USA}\\
{\em \small $(2)$ Instituut voor Theoretische Fysica, Universiteit Utrecht,}\\ 
{\em \small Princetonplein 5, 3584 CC Utrecht, The Netherlands}
}
\date{\today}
\maketitle

\begin{abstract}
We establish the equivalence of a recently introduced discrete model of
the hydrophobic interaction, as well as its extension to continuous
state variables, with the Ising model in a magnetic field with
temperature-dependent strength.  In order to capture the effect of
symmetries of the solvent particles we introduce a generalized
multi-state model. We solve this model -- which is not of the Ising
type -- exactly in one dimension. Our findings suggest that a small
increase in symmetry decreases the amplitude of the solvent-mediated
part of the potential of mean force between solute particles and
enhances the solubility in a very simple fashion. High symmetry
decreases also the range of the attractive potential. This weakening of
the hydrophobic effect observed in the model is in agreement with the
notion that the effect is entropic in origin.
\end{abstract}

\newpage

\section{Introduction}

The hydrophobic interaction is a property of aqueous solutions
which has a long and multi-faceted history of study \cite{Kauz59,BenN80}. 
We focus here on one particular issue, viz. the solvent-mediated interaction
between solute particles. The balance between an energetically favorable
accommodation of a solute particle between solvent particles and the 
entropically unfavorable necessity to place these solvent molecules in some 
particular orientation leads to an increase in free energy which is believed 
to be the basic mechanism of the hydrophobic effect. Since this
increase of free energy becomes weaker with smaller separation of solute
particles an effective solvent-mediated {\em hydrophobic attraction} arises
which is superimposed on and may even dominate the direct interaction
between solute particles.

This interplay between favorable energy and unfavorable entropy has been 
illustrated in an idealized lattice model where each solvent particle is allowed 
to be in one of $q$ discrete states (orientations) \cite{Kolo99}. There is a
nearest neighbor interaction between solvent particles with energy $w$ if both 
particles are in one particular state and $u$ (with $u>w$) otherwise.
A solute particle may be accommodated in the interstitial site between 
solvent particles, provided both solvent particles are in that special state,
which is energetically favorable by the amount $u-w$. Calculation of the
solvent-mediated part of the potential of mean force between solute particles
and of the solubility (done exactly in one dimension \cite{Kolo99} and 
numerically in two and three dimensions \cite{Bark00}) has revealed in all 
space dimensions an inverse relationship between the magnitude of the attractive 
force and its range. Furthermore, the solubility of the hydrophobe decreases 
with increasing temperature. In one dimension the decay of the potential of mean
force with distance is monotonic, while in two and three dimensions there
is an oscillatory modulation.

Here we first review and extend this basic model to continuous state variables,
thus being somewhat more realistic with regard to the interpretation of the 
molecular states as orientations (Sec. II) and show that both the original 
discrete model and the continuous model are equivalent to the ferromagnetic 
Ising model with a magnetic field with a temperature-dependent strength (Sec. III).
Since this mapping retains the 
local structure of the interactions, all the known results about the Ising model 
can be translated directly into properties of the hydrophobic model. Finally, 
we model symmetries of the solvent particles in a simple fashion by allowing
for the accommodation of solute particles between solvent molecules that are
in any $n$ rather than just one 
energetically favorable pair of special states (Sec. IV). The resulting 
$(n+1)$-state model is not equivalent to an Ising model, but may still be 
solved exactly in one dimension. These exact results are discussed in the 
concluding Sec. V.

\section{Model of the hydrophobic attraction with discrete and continuous
state variables}

In the model of Refs. \cite{Kolo99,Bark00} each solvent molecule is placed on
a $d$-dimensional lattice and can be found in $q$ different states. These
states may be thought of representing orientations with respect to some 
arbitrarily chosen reference axis. There is a nearest neighbor interaction
which favors  pairs of molecules which are both in one particular state.
We shall refer to this state as the special state, or state 1. Thus the 
interaction energy between molecules in states $i,j$ has the form 
\bel{1-1}
w_{ij} = \left\{ \ba{ll} w, & \mbox{ when $i=j=1$} \\
                         u, & \mbox{ otherwise} \ea   \right.
\ee
with
\bel{1-2}
u > w.
\ee   
Without loss of generality we shall set the global reference energy $u=0$
and therefore consider the case $w<0$. The energy of interaction of an
interstitial solute molecule with its neighbors is taken to be $v$. This
quantity, however, does not enter the expression for the solvent-mediated
part of the potential of mean force; only the solubility depends on $v$
\cite{Prat77}.

We calculate the potential of mean force $W(r)$ on the basis of the 
potential-distribution theorem \cite{Wido63}. We define $P_{11}$ as the 
probability of finding a given neighboring pair of molecules in the special 
state 1 and $P(r)$ as the probability of finding two such pairs at a distance 
$r$, measured in lattice units. Following \cite{Kolo99} one then finds
\bel{1-3}
W(r) = -kT \ln{[P(r)/P_{11}^2]}
\ee
at temperature $T$.

As solubility we define the dimensionless ratio 
\bel{1-4}
\Sigma = \rho_{\rm soln}/\rho_{\rm gas}
\ee
where $\rho_{\rm soln}$ is the number density of the solute in the solution
and $\rho_{\rm gas}$ is its number density in an ideal gaseous phase in osmotic 
equilibrium with the solution. For small solubility (such that the saturated 
solution is very dilute) this quantity is the Ostwald absorption coefficient. 
Within the lattice model one finds \cite{Kolo99}
\bel{1-5}
\Sigma = P_{11} \mbox{e}^{-v/kT}.
\ee

The equilibrium properties of the model are encoded in the partition function
\bel{1-6}
Z_L = \sum_{C} \mbox{e}^{-\beta E(C)}
\ee
where $L$ is the number of lattice sites, $\beta = 1/(kT)$, and the sum
is over all configurations $C$ the system may be found in. The nearest 
neighbor sum
\bel{1-7}
E(C) = w \sum_{<i,j>} \delta_{s_i,1}\delta_{s_j,1}
\ee
is the total energy of a configuration $C$ of state variables 
$s_i = 1, \dots, q$.

{}For the actual computation of the quantities $P_{11}$ and $P(r)$ one uses the
standard transfer matrix formulation of statistical mechanics models; see e.g. 
\cite{Baxt82}. In one dimension the transfer matrix is the $q\times q$ matrix
\bel{1-8}
{\bf V} = \left( \begin{array}{cccc}
b & 1 & \dots & 1 \\ 
1 & 1 & \dots & 1 \\ 
: & : &       & : \\ 
1 & 1 & \dots & 1
\end{array} \right)
\ee
which has 
\bel{1-9}
b = \mbox{e}^{-w/kT} > 1
\ee
at the (1,1) position and 1 everywhere else. The partition function becomes
\bel{1-10}
Z_L = \mbox{trace } {\bf V}^L.
\ee

In order to calculate thermal averages we also introduce the projectors
${\bf S}(k)=\ket{k}\bra{k}$ on states $k$. These are the diagonal matrices with $1$ 
at the diagonal element $k$ and 0 elsewhere. This yields the formal expressions
\bea
\label{1-11}
P_{11} & = & \frac{1}{Z_L} \mbox{trace } [{\bf S}(1){\bf V} {\bf S}(1) 
{\bf V}^{L-1}] = \frac{b}{Z_L}\bra{1} {\bf V}^{L-1} \ket{1} \\
P(r) & = & \frac{1}{Z_L} \mbox{trace } [{\bf S}(1){\bf V}{\bf S}(1) 
{\bf V}^{r-1}) {\bf S}(1){\bf V}{\bf S}(1) {\bf V}^{L-r-1}] \nonumber \\
\label{1-12}
     & = & \frac{b^2}{Z_L} \bra{1} \bf{V}^{r-1} \ket{1} \bra{1} 
\bf{V}^{L-r-1} \ket{1}
\eea
In the thermodynamic limit $L \to \infty$ the leading contribution to the
partition function comes from the largest eigenvalue $\lambda$ of the transfer
matrix. Anticipating the finite gap in the spectrum of ${\bf V}$ we may 
therefore write
\bea
\label{1-13}
P_{11} & = &  \frac{b}{\lambda} \rho_1 \\
\label{1-14}
P(r) & = & \frac{b^2}{\lambda^2} Q_{11}(r)
\eea
where $\rho_1$ is the probability of finding a molecule in state 1 and the
correlation function 
\bel{1-13a}
Q_{11}(r) = \lim_{L\to\infty} \frac{1}{Z_L} \mbox{trace } 
[{\bf S}(1) {\bf V}^{r-1} {\bf S}(1) {\bf V}^{L-r+1}]
\ee
is the probability of finding molecules at both sites $(i,i+r-1)$ in state 1, 
irrespective of the configuration of other molecules.

In higher dimensions the transfer matrix is constructed analogously. On a 
hypercubic lattice in $d$-dimensions the transfer matrix can be written as a 
product of two non-commuting matrices ${\bf V} = {\bf V}^{(2)} {\bf V}^{(1)}$. 
The diagonal matrix
\bel{1-15}
{\bf V}^{(1)} = \mbox{e}^{-\beta {\bf E}}
\ee
gives the Boltzmann weight of a configuration in a $(d-1)$-dimensional
hyperplane with
\bel{1-16}
{\bf E} = w \sum_{<i,j>} {\bf '} {\bf S}_i(1) {\bf S}_j(1).
\ee
The projectors ${\bf S}_i$ act as unit matrices on all sites
except $i$. The prime on the sum marks the restriction of the summation
to the hyperplane of states on which the transfer matrix acts.
The non-diagonal part
\bel{1-17}
{\bf V}^{(2)} = \prod_{i} {\bf V}_i
\ee
is the tensor product of the one-dimensional transfer matrix over all sites
of the hyperplane. Thermal averages are calculated by inserting projectors
${\bf S}_i(k)$ at the appropriate positions in ${\bf V}$ and in the trace 
over ${\bf V}^L$.

In the previous discussion the states that the molecule may be found in were 
chosen to belong to a discrete set. Since the model is not designed to be 
realistic for any particular solvent such an artificial representation of the 
orientation of a molecule is not really a cause for concern. Nevertheless, for 
a proper interpretation of these states as orientations it is useful to extend 
the model to continuous state variables by the following limit procedure. One
imagines there to be $n$ distinct states $1, \dots, n$ which have mutual 
interaction energy $w$. Thus there are $n^2$ pairs of states with energy $w$ 
and $q^2-n^2$ pairs which have interaction energy $u=0$. Hence we shall refer 
to this model as the $n^2$-model. These $n$ states may be interpreted as 
labeling infinitesimal solid angle segments on a sphere surrounding the solvent
molecule (see Fig.~\ref{F-1} for a two-dimensional representation). Not all of
these segments have to be in close proximity on the sphere; they may belong
to distinct blocks which define those areas on the surface of the molecule
between which a solute particle may be accommodated. Taking the limit
$n,q \to \infty$ with the ratio $\tilde{q} = q/n$ fixed, one obtains a continuous
version of the model where the range of states $[0,4\pi/\tilde{q}]$ represents
the total amount of solid angle on the surfaces of each of the two 
neighboring solvent molecules
between which solute molecules may be found. Parts of that interval may then 
refer to distinct areas on the surface of the solvent particle.

{}For fixed $n$ and $q$ the one-dimensional transfer matrix has a block form
similar to (\ref{1-8}), but where instead of the matrix element $b$ one
has an $n \times n$ matrix with all elements equal to $b$. All other matrix
elements are equal to 1. As will be shown in the next section, the $n^2$-model 
and the original $1^2$-model are equivalent, thus proving that the earlier
restriction to discrete states did not constitute an additional 
oversimplification. In the interpretation of the $1^2$-model as a model with a 
continuous degree of freedom the special state 1 simply represents a fraction 
$4\pi/q$ of solid angle (oriented in some direction) rather than a discrete 
single direction. Notice, however, that the analogue of the quantity $P_{11}$ 
is the double sum 
\bel{1-18}
P_{11} \to \sum_{i,j=1}^n P_{i,j}
\ee
which turns into an integral in the limit $n,q \to \infty$.
The pair correlation function $P(r)$ has to be redefined accordingly.
With the proper redefinitions all results of the $1^2$-model may be 
interpreted in terms of the continuum limit of the $n^2$-model, i.e.,
with the integer $q$ being replaced by $\tilde{q}=q/n$, 
and $1<\tilde{q}<\infty$ considered a real number.

\section{Equivalence to the Ising model}

In the $n^2$-model all special states are indistinguishable, as are the 
non-special states. Hence it is important to know only to which group
of states the state of a molecule belongs, but not to which particular
state within each group. Any joint probability involving some state
$k$ depends only on whether $k$ belongs to the group of special states
$(k \in \{1, \dots, n\})$ or not ($k \in \{n+1, \dots, q\}$). Hence the 
$n^2$-model is equivalent to a 2-state model, i.e., to an Ising model.
We may identify the special states with an Ising spin $s_i=+1$ whereas 
the other states may be collectively identified with $s_i=-1$. 

In order to work out the equivalence in detail we note that
the partition function (\ref{1-6}) of the original $1^2$-model may be cast 
in the form of a double sum
\bel{2-1}
\sum_C \mbox{e}^{-\beta E(C)} = \sum_{C_1} \sum_{C'(C_1)} 
\mbox{e}^{-\beta E(C)}.
\ee
The first sum runs over all fixed configurations $C_1$ of $N_1$ molecules in 
state 1; the second sum then includes all other configurations $C'$ with the 
configuration $C_1$ specified by the first sum. Since the energy $E(C)=E(C_1)$
depends only on the positions of molecules in state 1, i.e., on $C_1$, the 
second sum can be evaluated trivially to give 
\bel{2-2}
\sum_{C'(C_1)} = (q-1)^{L-N_1}.
\ee
This originates in the degeneracy of states with a fixed number $N_1$ of 
molecules in state 1 in the configuration $C_1$, and we obtain
\bel{2-3}
Z_L = \sum_{C_1} \mbox{e}^{-\beta E(C_1)} (q-1)^{L-N_1} = 
\sum_{C_1} \mbox{e}^{-\beta \tilde{E}(C_1)}
\ee
with a new energy function
\bel{2-4}
\tilde{E} = E + kT \ln{(q-1)} \sum_{i} (1 - \delta_{s_i,1}).
\ee

In those terms the partition function (\ref{2-3}) of the model for hydrophobic 
attraction turns into the partition of a ferromagnetic Ising model in a 
magnetic field with energy
\bel{2-5}
\tilde{E} = - J \sum_{<i,j>} s_i s_j - h \sum_{i} s_i - const.
\ee
The Ising interaction parameters are given by the relations
\bea
\label{2-6}
J & = & -\frac{w}{4} \\
\label{2-7}
h & = & -\frac{cw}{4} - \frac{1}{2} kT \ln{(q-1)},
\eea
with $c$ the coordination number of the lattice. This enters since part of the 
contribution of the magnetic field strength is contained in the nearest
neighbor sum (\ref{1-7}). If some lattice sites have a coordination number
different from $c$ (e.g. at the boundary in case of non-periodic boundary
conditions), additional local magnetic fields would contribute to the Ising 
energy. The temperature dependence of the entropic contribution $kT\ln{(q-1)}/2$ 
to the field arises since the number of unfavorable states does not depend on 
temperature and hence the factor $kT$ must cancel in the partition function. 
The constant term in (\ref{2-5}) does not affect the thermodynamical properties 
of the system and may be dropped. Notice that $q$ is a discrete variable and 
therefore within this formulation of the model $h$ changes discontinuously as a 
function of q.

A similar resummation may be performed for the partition function of the
$n^2$-model by first fixing a configuration of favorable states with $N_f$
molecules (which yields a factor $(q-n)^{L-N_f}$ from the summation
over the remaining degenerate configurations) and then counting the
degeneracy of all such favorable states. For each given $N_1=N_f$ this yields 
another factor $n^{N_1}$. Thus we are left with an Ising partition function 
with a magnetic field
\bel{2-8}
h' = -\frac{cw}{4} - \frac{1}{2} kT \,\ln{(q/n-1)}.
\ee
The continuum limit of the state variable may now be taken and yields
an Ising partition function identical to that derived from the original 
$1^2$-model but where now $q$ is is replaced by $\tilde{q}=q/n$ and treated
as a continuous variable in the range $1 < \tilde{q} < \infty$. Hence the 
$n^2$-models are all equivalent to each other and to the Ising model in a 
magnetic field given by (\ref{2-8}), i.e., by the replacement $q \to q/n$. 
Without loss of generality one may therefore discuss only the $1^2$-model.

On the level of the transfer matrix description the equivalence may be shown
by a similarity transformation ${\bf \tilde{V}} = {\bf Y} {\bf V} {\bf Y}^{-1}$. 
Because of the cyclic property of the trace the partition function is invariant
under any such transformation. For the $1^2$-model we define
\bel{2-9}
\omega = \mbox{e}^{2\pi i/(q-1)}
\ee
The matrix
\bel{2-10}
{\bf Y} = \frac{1}{\sqrt{q-1}} \left( \begin{array}{cccccc}
\sqrt{q-1} & 0 &       0       &       0        & \dots &      0           \\ 
       0   & 1 &       1       &       1        & \dots &      1           \\ 
       0   & 1 & \omega^{-1}   & \omega^{-2}    & \dots & \omega^{-(q-2)}  \\
       0   & 1 & \omega^{-2}   & \omega^{-4}    & \dots & \omega^{-2(q-2)} \\
       :   & : &       :       &       :        &       &      :           \\ 
       0   & 1 &\omega^{-(q-2)}&\omega^{-2(q-2)}& \dots & \omega^{-(q-2)^2}
\end{array} \right)
\ee
has as its inverse the transposed matrix, $Y^{-1} = Y^T$. The transformation
leaves the projector ${\bf S}(1)$ invariant and brings ${\bf V}$ into a block 
diagonal form with the $2\times 2$ matrix 
\bel{2-11}
{\bf W} = \left( \ba{cc}
                      b     & \sqrt{q-1} \\
                 \sqrt{q-1} &     q-1 \ea \right)
\ee
in the upper left diagonal and 0 everywhere else. Therefore the trace reduces
to a trace over the two-dimensional subspace on which ${\bf \tilde{V}}$ acts in
a nontrivial way, i.e., $\mbox{trace } {\bf V}^L = \mbox{trace }{\bf \tilde{V}}^L
 = \mbox{trace } {\bf W}^L$. In ${\bf W}$ one recognizes the transfer 
matrix of the Ising model and hence the equivalence is recovered. In higher 
dimensions the transformation ${\bf Y}$ is applied to each lattice site, thus
reducing the full model to a two-state Ising model with parameters as defined
above.

Varying the temperature $T$ corresponds to moving along a curve in the
field-temperature plane of the Ising model which starts at
$T=0, h=-cw/4 >0$ (all molecules are in the favorable state) and crosses the 
line $h=0$ at a point defined by the condition
\bel{2-12}
- w = \frac{2}{c} kT \, \ln{(q-1)}
\ee
or, equivalently, $b^{c/2}=q-1$ (Fig~\ref{F-2}). In two and three dimensions
the Ising model has a phase transition of second order at finite temperature
along this zero-field line. Above the critical point the probability of 
finding a molecule in the favorable state is 1/2, below the critical point 
one has spontaneous symmetry breaking between a ``favorable'' and an
``unfavorable'' phase. On a $d$-dimensional square lattice ($c=2d$) some
values of $q$ at the critical point are given by \cite{FYWu82}
\bel{2-13}
q_c = 1 + \mbox{e}^{-dw/kT_c} = \left\{ \ba{ll}
      \infty          & d=1       \mbox{ (exact)}\\
      18 + 12\sqrt{2} & d=2       \mbox{ (exact)} \\
      \approx 27      & d=3       \mbox{ (numerical)}\\
      1 + \mbox{e}^2  & d =\infty \mbox{ (mean field)} \ea \right.
\ee
At the critical point the range of the potential of mean force would diverge
if the model were applicable at such low values of $q$. At large distances the 
decay would be algebraic and proportional to the decay of the
magnetization correlation function.

{}For larger values of $q$ (which seem more realistic) the zero-field condition 
corresponds to subcritical temperatures. Hence, following the curve defined by 
varying the temperature of the hydrophobic model one expects a jump 
discontinuity in the
probability of finding a favorable state from a value greater than 1/2 to
a value less than 1/2 at $h=0$. The strength of the discontinuity increases with
increasing $q$. The regime which appears to be relevant for the hydrophobic
effect \cite{Kolo99,Bark00} corresponds to $h<0$. For all coordination
numbers ($c=2$ being the smallest possible) this is consistent with the
requirement that the change of free energy $\Delta F = w + kT \ln{(q-1)}$ 
arising from the accommodation of a solute particle be positive.

\section{Symmetry effects}  

Implicit in the construction of the $1^2$-model is the view that 
the surface structure of the solvent molecule is made up of certain regions 
between which a solute molecule may be accommodated. No provision is made for 
more structured solvent molecules which have chemically different
surface regions. One could imagine $n$ non-identical regions on each solvent
molecule such that each identical pair of regions allows for the accommodation 
of a solute particle between two solvent molecules, but the accommodation 
between distinct pairs would be energetically so unfavorable that it may be 
considered forbidden. In this section we extend the model to allow for $n$ 
pairs of orientations between which a solute particle may be accommodated
(Fig.~\ref{F-1}). 
We denote these special states by $1, \dots, n$, with $1 \leq n \leq q$ and 
assign non-positive interaction energies
\bel{3-1}
w_{ij} = \left\{ \ba{ll} w, & \mbox{ when $1\leq i=j \leq n$} \\
                         0, & \mbox{ otherwise} \ea   \right.
\ee

We shall refer to this model as the $n$-model. With regard to solubility 
a solvent molecule of type $n$ described by these interactions would have a 
higher symmetry than a molecule described by the original $1$-model with the
same value of $q$, since there are more, but mutually exclusive pairs
of favorable orientations. The limiting case $n=q$ corresponds to the
zero-field, $q$-state Potts model \cite{FYWu82}. We stress that we define this 
model in the spirit of Refs. \cite{Kolo99,Bark00}, i.e., not with the intention 
of describing a particular real solvent, but with the aim of exhibiting what may 
be found to emerge as generic behavior resulting from 
an increase in symmetry. Since some of the important features of the $1$-model 
were already seen in the simple one-dimensional model, we restrict our analysis 
of the $n$-model also to the one-dimensional case. This can be solved
exactly by diagonalizing the transfer matrix. A reader not interested
in the details of the diagonalization may skip the following 
more technical subsection.

\subsection{Exact solution of the one-dimensional model}

The $q\times q$ transfer matrix ${\bf V}$ of the $n$-model has 
$b$ for its first $n$ diagonal elements and 1 everywhere else. By
a transformation similar to (\ref{2-10}) -- with the 1 at the (1,1)-position
replaced by an $n \times n$ unit matrix and the quantity $q-1$ in the
normalization factor and in $\omega$ replaced by $(q-n)$ -- the transfer matrix
becomes block diagonal with
\bel{3-2}
{\bf W} = \left( \begin{array}{cccccc}
       b   &     1      &   \dots  &     1      & \sqrt{q-n}  \\ 
       1   &     b      &   \dots  &     1      & \sqrt{q-n}  \\ 
       :   &     :      &                :      &       :     \\ 
       1   &     1      &   \dots  &     b      & \sqrt{q-n}  \\
\sqrt{q-n} & \sqrt{q-n} &   \dots  & \sqrt{q-n} &   q-n
\end{array} \right)
\ee
in the upper left corner and 0 elsewhere. Hence the model is equivalent to an
$(n+1)$-state model, which is, however, not equivalent to some Ising or
Potts model. Nevertheless the degree of degeneracy is high and therefore
it is not difficult to guess the eigenvectors of ${\bf W}$ and hence of 
${\bf V}$ by a second transformation, again analogous to ${\bf Y}$, but with 
the first two columns replaced by matrix elements of the form $Y_{11} = y_+/z_+$,
$Y_{k1} = 1/z_+$ and $Y_{12} = y_-/z_+$, $Y_{k2} = 1/z_-$ for all 
$2 \leq k \leq n+1$, and with the quantity $q-1$ replaced by $n$. Since the
columns of the transformation matrix are eigenvectors of ${\bf W}$
the variables 
\bel{3-4}
y_{\pm} = \frac{1}{2\sqrt{q-n}}\left(q+1-2n-b\pm 
\sqrt{(b-q-1)^2+4n(b-1)}\right).
\ee
are determined by the two solutions of a quadratic equation. The constants 
\bel{3-5}
z_\pm^2 = n+y_\pm^2.
\ee
are normalization factors for the eigenvectors.

The full transformation which diagonalizes ${\bf V}$ may then be constructed
and all quantities of interest may be calculated. One finds four distinct
eigenvalues which take the values
$b-1$ (which is ($n-1$)-fold degenerate), 0 (($q-n-1$)-fold degenerate),
and
\bel{3-3}
\lambda_{\pm} = \frac{1}{2} \left(b+q-1 \pm \sqrt{(b-q-1)^2+4n(b-1)}\right).
\ee
The largest eigenvalue is $\lambda_+$. 
Before proceeding further we give the asymptotic expansion up to first order
in $1/q$ which is necessary for the study of the hydrophobic
regime of large values of $q\gg b$. One finds:
\bea
\label{3-foa}
\lambda_+ & = & q + (b-1)n/q \\
\label{3-fob}
\lambda_- & = & (b-1)(1-n/q) \\
\label{3-foc}
z_+^2 & = & q-2(b-1)(1-n/q) \\
\label{3-fod}
z_-^2 & = & \left[q+2(b-1)n/q\right]n/(q-n).
\eea

The exact components $\lambda^{(k)}$ of the eigenvectors $\ket{\lambda}$ 
belonging to the eigenvalues $\lambda$ are given as follows:\\[4mm]
\underline{$\lambda=\lambda_{\pm}$:}\\
\bea
\lambda^{(k)} & = & 1/z_\pm \quad (1 \leq k \leq n) \\
\lambda^{(k)} & = & y_\pm/(z_\pm\sqrt{q-n}) \quad (n+1 \leq k \leq q) 
\eea
\underline{$\lambda=b-1$:} ($p=1,\dots,n-1$; $\omega_1 = \mbox{e}^{2\pi i/n}$)\\
\bea
\lambda_p^{(k)} & = & \omega_1^{p(k-1)}/\sqrt{n} \quad (1 \leq k \leq n) \\
\lambda_p^{(k)} & = & 0 \quad (n+1 \leq k \leq q )
\eea
\underline{$\lambda=0$:} ($p=1,\dots,q-n-1$;
$\omega_2 = \mbox{e}^{2\pi i/(q-n)}$)\\
\bea
\lambda_p^{(k)} & = & 0 \quad (1 \leq k \leq n) \\
\lambda_p^{(k)} & = & \omega_2^{p(k-n-1)}/\sqrt{q-n} \quad (n+1 \leq k \leq q)
\eea

The matrix elements needed for the calculation of average values presented
below are then given by
\be
\bra{k} {\bf V}^n \ket{k'} = \left\{ \ba{l}
\frac{\lambda_+^n}{z_+^2} + 
(b-1)^n\left(\delta_{k,k'}-\frac{1}{n}\right)
+ \frac{\lambda_-^n}{z_-^2} \quad (1\leq k,k'\leq n) \\
\frac{\lambda_+^ny_+}{z_+^2\sqrt{q-n}} + \frac{\lambda_-^ny_-}{z_-^2\sqrt{q-n}} 
\quad (1\leq k\leq n;\,n+1 \leq k'\leq q) \\
\frac{\lambda_+^ny_+^2}{z_+^2(q-n)} + \frac{\lambda_-^ny_-^2}{z_-^2(q-n)} 
\quad (n+1 \leq k,k'\leq q) \ea \right. .
\ee
These expressions are obtained by inserting a complete set of eigenstates
of the transfer matrix,
\be
\bra{k} {\bf V}^n \ket{k'} = \sum_\lambda 
\langle\,k\,\ket{\lambda}\bra{\lambda} {\bf V}^n \ket{k'},
\ee
and using the components $\lambda^{(k)} = \bra{k}\,\lambda\,\rangle$ and their
complex conjugates $\lambda^{(k)\ast} = \bra{\lambda}\,k\,\rangle$ given
above.

\subsection{Potential of mean force and solubility}

To study the properties of the model we consider first
\bel{3-6}
P(k) = \lim_{L\to\infty} \mbox{trace } ({\bf S}(k) {\bf V}^L) / Z_L
= \lim_{L\to\infty} \bra{k} {\bf V}^L \ket{k}/Z_L
\ee
which is the probability of finding a molecule in state $k$. Using the
matrix elements calculated in the above one finds
\bel{3-7}
P(k) = \left\{ \ba{ll} 1/z_+^2 & 1 \leq k \leq n \mbox{ (special states)}\\
                       y_+/(z_+^2(q-n)) & n+1 \leq k \leq q \ea \right.
\ee
At the temperature defined by
\bel{3-8}
- w = kT \, \ln{(q-2n+1)}
\ee
the probability 
\bel{3-9}
P_s = \sum_{k=1}^n P(k) = n/z_+^2
\ee 
of finding a molecule in any one of the special states
equals 1/2. This is the analogue of the zero field condition (\ref{2-12}) in the 
1-model in one dimension and suggests that the condition $q-2n+1 > b$ limits 
the parameter range of the $n$-model relevant for modeling the hydrophobic 
effect.

The quantity entering the solubility is the nearest-neighbor correlation
function
\bel{3-10}
P_{11} =\lim_{L\to\infty} \sum_{k=1}^n \mbox{trace } 
({\bf S}(k){\bf V}{\bf S}(k){\bf V}^{L-1})/Z_L = \frac{b}{\lambda_+} P_s.
\ee
{}For the computation of the potential of mean force we also need
\bea
P(r) & = & \lim_{L\to\infty} \sum_{k=1}^n \sum_{l=1}^n \mbox{trace } 
           [{\bf S}(k){\bf V}{\bf S}(k){\bf V}^{r-1}{\bf S}(l){\bf V}{\bf S}(l)
           {\bf V}^{L-r-1}]/Z_L \nonumber \\
     & = & \frac{b^2}{\lambda_+^2} \lim_{L\to\infty} \sum_{k=1}^n \sum_{l=1}^n 
     \mbox{trace } [P(k){\bf V}^{r-1} P(l){\bf V}^{L-r-1}]/Z_{L-2}\nonumber \\
\label{3-11}
     & = & \frac{b^2n^2}{\lambda_+^2z_+^4}\left[ 1+\frac{z_+^2}{z_-^2} 
     \left(\frac{\lambda_-}{\lambda_+}\right)^{r-1}\right].
\eea
The potential of mean force therefore takes the form
\bel{3-12}
W(r) = - kT \ln{\left[1+A\mbox{e}^{-(r-1)/\xi}\right]}
\ee
with the amplitude
\bel{3-13}
A = \frac{z_+^2}{z_-^2}
\ee
and the localization length
\bel{3-14}
\xi = \left(\ln{\frac{\lambda_+}{\lambda_-}}\right)^{-1}.
\ee
We remark that the expression of the potential of mean force does not involve
the eigenvalue and eigenvectors of the the $(n-1)$-fold degenerate eigenvalue
$b-1$ characteristic for the $n$-model. The eigenvalues $\lambda_\pm$ and the
quantities $z_\pm$, however, do depend on $n$.

The amplitude $A$ is a monotonically increasing function of temperature, with 
$A \sim (q/n-1)n^2/b^2$ as $T\to 0$ and $A=q/n - 1$ at infinite temperature. 
The localization length vanishes in both these limiting cases
and has a maximum at $b=q+1$, independently of $n$. For this temperature
we find in terms of the ratio $\nu = n/q$
\bea
\label{3-15}
A & = & \frac{1-\sqrt{\nu}}{1+\sqrt{\nu}} \\
\label{3-16}
\xi & = & \left[\ln{\frac{1+\sqrt{\nu}}{1-\sqrt{\nu}}}\right]^{-1} 
\approx 1/\sqrt{4\nu} \\
\label{3-17}
\Sigma & = & \frac{1}{2} \nu b \mbox{e}^{-v/(kT)}.
\eea
{}For large $q$ the maximal localization length grows proportionally to the 
square root of $q$, but decreases with increasing $n$. As a function of $n$
the amplitude of the potential of mean force also decreases. The solubility
is enhanced by a factor of $n$.

In analogy to the original 1-model one expects the regime most relevant
to the hydrophobic effect to correspond to $q \gg b$. Using again $\nu = n/q$
and expanding the relevant
quantities $\lambda_\pm$, $z_\pm^2$ up to first order in $1/q$, (\ref{3-foa}) -
(\ref{3-fod}) yield
\bea
\label{3-18}
A & = & \frac{1-\nu}{\nu} \\
\label{3-19}
\xi & = & \left[\ln{\frac{q}{(b-1)(1-\nu)}}\right]^{-1} \\
\label{3-20}
\Sigma & = & \frac{b \nu}{q}\mbox{e}^{-v/(kT)}.
\eea
The inverse localization length may be written as a sum of two terms,
\bel{3-21}
\xi^{-1} = \ln{\frac{b}{q-1}} + \ln{\frac{1}{1-\nu}},
\ee
the second of which is positive and describes the reduction of $\xi$ as a 
function of $n$. As in the previous special case the amplitude decreases
with increasing $n$, while the solubility increases by a factor of $n$.

At sufficiently large distance the potential of mean force decays exponentially
\bel{3-22}
W(r) \sim -kT A \mbox{e}^{-(r-1)/\xi}
\ee
{}For small $n$ and large $q$ the localization length $\xi$ depends on
$n$ only very weakly. The amplitude is reduced by a factor of $n$ while
the solubility is increased by the same factor. Hence one finds for two
different systems the relations
\be
\label{3-23}
\frac{W(r)}{W'(r)} = \frac{\nu'}{\nu} \quad \quad \mbox{ for $\xi=\xi'$}
\ee
\be
\label{3-24}
\frac{\Sigma}{\Sigma'}= \frac{\nu q'}{\nu' q}  \quad \quad 
\mbox{ for $T(v'+w')=T'(v+w)$}
\ee
\be
\label{3-25}
1/\xi - 1/\xi' = \ln{(q/q')} \quad \quad  \mbox{ for $Tw'=T'w$}.
\ee
One can analyze in a similar fashion the effect of symmetry within the
$1^2$-model where higher symmetry corresponds to a larger solid angle of
accommodation and hence to a decrease in $q$. This corresponds to setting
$\nu=1/q$ in Eqs. (\ref{3-23}) - (\ref{3-25}).

\section{Concluding remarks}

The equivalence of the hydrophobic interaction model to the Ising model
in a magnetic field with temperature-dependent strength allows for
the interpretation of the solubility, the potential of mean force and
other quantities of interest in terms of the well-known properties of the
Ising model. For realistic values for water in the hydrophobic model,
the strength of the magnetic field in the corresponding Ising model is
significant, and consequently the model is far away from the Ising
critical point.  We can however not rule out that some real system {\em
is} appropriately described by an Ising model, in which case a
solvation effect associated with the corresponding short range
interaction would change dramatically at the critical point.
Indeed, the interpretation of $q$ as a fraction of the
surface area between which solute particles may be accommodated suggests
lower realistic values of $q$ than previously assumed and hence the zero-field
condition for criticality might be satisfied if such a system exists.
A more tangible consequence of a lower value of $q$ (but not as low
as required for the second order phase transition) would be to bring the 
minimum of the solubility as a function of temperature to larger 
and more realistic values, thus indicating that the mechanism by which this
minimum occurs in a real system is similar to that described by the model.

The most striking features of the original hydrophobic interaction model,
the inverse relationship between amplitude and range of the potential of
mean force and the decrease of solubility with increasing temperature
in reasonable temperature ranges, are present also in the more symmetric
$n$-model. However, the amplitude of the potential decreases with increasing 
degree of symmetry. This is in agreement with the notion that the origin of the 
hydrophobic effect is entropic. With increasing degree 
of degeneracy (and hence symmetry) the entropy decrease associated with the
requirement to orient molecules in a particular direction is less
and hence also the balance in the free energy difference is reduced.
{}For solvents differing in their degree of symmetry the $n$-model predicts
an inverse relationship (\ref{3-23}) also between the ratios of the
potential of mean force (at sufficiently large distance and temperatures
at which the localization length are equal) and the ratios of solubilities
(at temperatures such that the exponential prefactors are equal).

\section*{Acknowledgments}
This work has been supported by the U.S. National Science Foundation and the
Cornell Center for Materials Research. G.M.S. would like to thank the 
Department of Chemistry and Chemical Biology, Cornell University for kind 
hospitality.

\newpage
\begin{figure}[tbp]
\begin{center}
\epsfxsize=0.4\textwidth
\epsffile{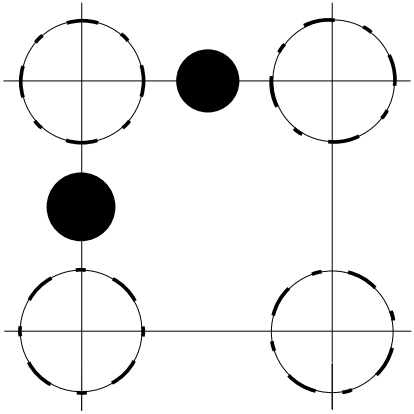}
\end{center}
\caption{Accommodation of solute particles (black disks) on the interstitial 
sites between solvent particles (circles) on a square lattice. Depending on the 
model the figure has different interpretations. (i) In the $n^2$-model with 
$q$ discrete orientations (states) the whole circle consists of $4q$ discrete 
points which consist of quadruples related to each other by a 90 degree 
rotation. Each quadruple defines one orientation and there
is always one quadruple of points which is located on the lattice axis.
The special states $1,\dots,n$ correspond to quadruples of points on the dark 
sections of the circumference of the circles. A solute particle may be 
accommodated whenever special points of two neighboring molecules are located 
on the lattice axis. (In the $1^2$-model of Ref. \protect\cite{Kolo99}
there would be only one such quadruple of points.) (ii) In the continuum limit 
the interpretation of the open circles in the figure is that if dark segments
of the circumference of neighboring 
circles intersect with the lattice axis then a solute particle may be 
accommodated. (iii) In the $n$-model defined below one distinguishes between 
$n$ different special states. Accommodation of solute particles is possible 
only between neighboring solvent molecules which are in the same special 
state corresponding to one particular of $n$ special quadruples of points. In 
the continuum version of the $n$-model this corresponds to distinguishing 
between different surface areas on the solvent molecule. E.g. for $n=2$ one 
would say that a solute particle could be accommodated if on both neighboring 
molecules
either the long or the short dark segment intersect with the lattice axis,
but not if for one molecule the long segments intersects and for the other
molecule the short segment. (This scenario would allow for accommodation
of a solute particle in the continuum version of the $n^2$-model.)}
\label{F-1}
\end{figure}

\begin{figure}[tbp]
\epsfxsize=0.7\textwidth
\epsffile{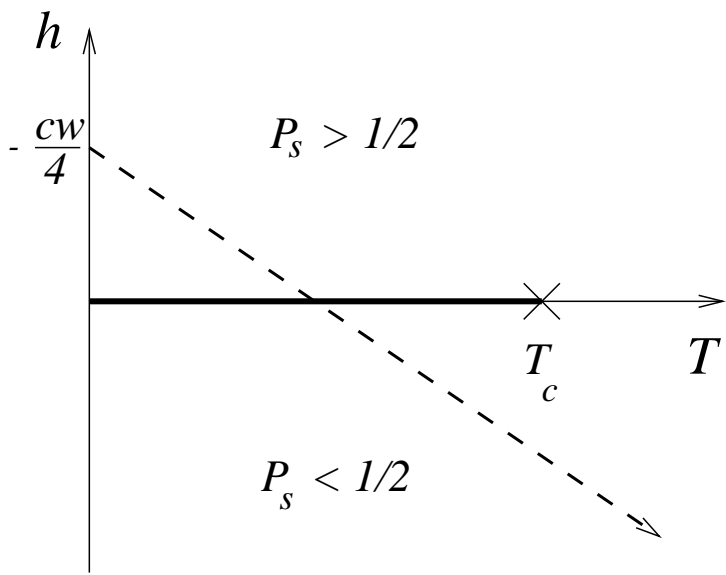}
\caption{Curve in the $h-T$ plane of the Ising model which corresponds to
varying $T$ in the hydrophobic model (dashed line). Larger $q$ leads to
a steeper (more negative) slope. The bold line at $h=0$ marks the
region of spontaneous symmetry breaking, ending in the critical point at $T_c$.
In the regime $h<0$ relevant for the hydrophobic effect the probability $P_s$ 
of finding a solvent molecule in the special state is less than 1/2.}
\label{F-2}
\end{figure}

\end{document}